\documentclass[prb,aps,showpacs,floatfix,twocolumn,10pt]{revtex4-1}

\pdfoutput=1

\usepackage[english]{babel}

\usepackage[utf8]{inputenc}

\usepackage{amsmath,amssymb,amstext}

\usepackage{amsthm}

\usepackage{txfonts}

\usepackage{lmodern}

\usepackage{xcolor,soul}

\usepackage{bm}

\usepackage{graphicx}

\usepackage{array}

\usepackage[caption=false]{subfig}

\usepackage[activate=normal]{pdfcprot} 

\usepackage{siunitx} 

\usepackage{hyphenat}

\newcommand{\vect}[1]{\bm{#1}}

\newcommand*\diff{\mathop{}\!\mathrm{d}}

\newcommand{\cosimSW}{Coral}

\newcommand{\sub}{subsimulator}
\newcommand{\Sub}{Subsimulator}

\newcommand{\guidelineCosim}{%
	Use co-simulation to construct full-system models from loosely coupled stand-alone models and modules.
}
\newcommand{\guidelineCosimshort}{%
	Couple models and tools using co-simulation.
}

\newcommand{\guidelinePowerbonds}{%
	Properly define and use high-level interfaces to guarantee interoperability of simulation models.
	Make use of power bonds to model the flow of energy between \sub{}s whenever possible.
	The use of SI units is highly advised.
	If other units are used, explicitly and clearly document so.
}
\newcommand{\guidelinePowerbondsshort}{%
	Use power bonds to model relevant energy transactions between \sub{}s whenever possible.
}

\newcommand{\guidelineErrorestimation}{%
	Use an error estimation method to assess and control co-simulation coupling errors, and guarantee the quality and validity of the simulation results.
}
\newcommand{\guidelineErrorestimationshort}{%
	Implement a co-simulation error estimation method.
}

\newcommand{\guidelineFMI}{%
	Use the Functional Mock-up Interface to ensure compatibility between different simulation tools and languages.
	Packaging \sub{}s as functional mock-up units makes them tool independent and re-usable.
}
\newcommand{\guidelineFMIshort}{%
	Use the Functional Mock-up Interface (FMI) to ensure interoperability and re-usability of \sub{}s.
}

\newcommand{\guidelineFUs}{%
	Use dedicated time independent submodules for unit conversions, coordinate transformations, signal algebra, and other generic operations between \sub{}s during a simulation, as well as to include (approximately) instantaneous phenomena.
	This ensures modularity and ease-of-use, and avoids unnecessary time delays.
}
\newcommand{\guidelineFUsshort}{%
	Consistently carry out generic signal operations between \sub{}s using dedicated time independent modules.
}

\newcommand{\guidelineReticulation}{%
	Try to obtain a good balance between modularity, complexity, accuracy, and numerical stability when splitting up a given system for co-simulation.
	Beware of time delays between \sub{}s, and consider that accuracy may suffer with increasing modularity.
	At the same time, try to provide a sufficient level of modularity to facilitate interoperability and re-usability of models.
}
\newcommand{\guidelineReticulationshort}{%
	Take great care when choosing a system reticulation, and try to balance modularity against complexity and numerical stability.
}

\newcommand{\guidelineTightcoupling}{%
	Avoid exposing tight couplings on the system level whenever possible in order to minimize coupling errors and avoid issues with numerical stability.
}
\newcommand{\guidelineTightcouplingshort}{%
	Try to avoid tight couplings on the system level whenever possible.
}

\newcommand{\guidelineCausality}{%
	For \sub{}s where the preferred causality--connectivity option is difficult or impossible to determine without prior knowledge of the connecting environment, the \sub{}s may be implemented as hybrid causality models to ensure compatibility with other \sub{}s in the co-simulation environment.
}
\newcommand{\guidelineCausalityshort}{%
	If the computational causality for a given \sub{} can not be determined a priori, implementation as a hybrid causality model is advised.
}

\newlength{\graphicswidth}
\newlength{\graphicswidthfull}
\theoremstyle{plain}
\newtheorem{guideline}{Guideline}

\theoremstyle{remark}
\newtheorem{remark}{Remark}[guideline]

\definecolor{orange}{rgb}{1,0.5,0}
\definecolor{gray}{gray}{0.5}

\begin{document}


\title{Distributed Co-Simulation of Maritime Systems and Operations}
\author{Severin~Sadjina, Stian~Skjong, Eilif~Pedersen, Vilmar {\AE}s{\o}y}
\affiliation{Department of Marine Technology, Norwegian University of Science and Technology, NO-7491 Trondheim, Norway}
\author{Lars~Tandle~Kyllingstad, Martin Rindar{\o}y, Dariusz Eirik Fathi, Vahid Hassani, Trond Johnsen, J{\o}rgen Bremnes Nielsen}
\affiliation{SINTEF Ocean, NO-7465 Trondheim, Norway}

\begin{abstract}
Here, we present the concept of an open virtual prototyping framework for maritime systems and operations that enables its users to develop re-usable component or subsystem models, and combine them in full-system simulations for prototyping, verification, training, and performance studies.
This framework consists of a set of guidelines for model coupling, high-level and low-level coupling interfaces to guarantee interoperability, a full-system simulation software, and example models and demonstrators.
We discuss the requirements for such a framework, address the challenges and the possibilities in fulfilling them, and aim to give a list of best practices for modular and efficient virtual prototyping and full-system simulation.
The context of our work is within maritime systems and operations, but the issues and solutions we present here are general enough to be of interest to a much broader audience, both industrial and scientific.
\end{abstract}

\maketitle



\section{Introduction}
\label{sec:introduction}

With operations that are becoming increasingly complex and demanding, volatile economic conditions, stricter environmental and safety regulations, and decreasing project lead times, simulation methods have become a key indicator of merit in early design phases within the maritime industry.
They allow for a quick exploration of the design space and help to advance concepts into certain directions.\cite{Harries2011}
Simulation methods also expose concept, interface, and safety flaws which in turn helps reduce risks and enhance operational performance and efficiency.

Traditional ship design is a sequential process:\cite{Evans1959}
Development is driven step-by-step and iteratively with each engineering discipline using its own set of tools that are rarely inter-operational.
This complicates system level analysis and verification significantly and obscures errors and issues until late in the design process when they are difficult and expensive to fix.
Because of this there is now an increasing interest in being able to fully integrate simulation techniques into the ship design process for prototyping, verification, training, and performance studies.\cite{Hassani2016}
Through the development of virtual prototypes it is possible to test the characteristics and dynamics of a proposed concept early on, and expose possible issues long before the integration and prototyping stages during which they usually surface.
This is especially important for the maritime industry which---unlike other industries such as automotive, aerospace, and railway---typically has to provide one-of-a-kind solutions.
Because of this, it is hard to learn lessons and advance with a traditional sequential process, when designs are unique and different each time around.

Virtual prototyping also holds the promise of substantially simplifying the search for an optimal design and helping to keep costs, risk factors, and environmental impacts low.
For a typical offshore supply vessel, for example, the power consumption of the on-board equipment is significant and will, to a large extend, determine the dimensions of the energy systems.
At the same time, there is an incentive to minimize these dimensions from economic and environmental standpoints in order to keep the entire vessel as small as possible.

\begin{figure}[htb]
	\includegraphics[width=\graphicswidth]{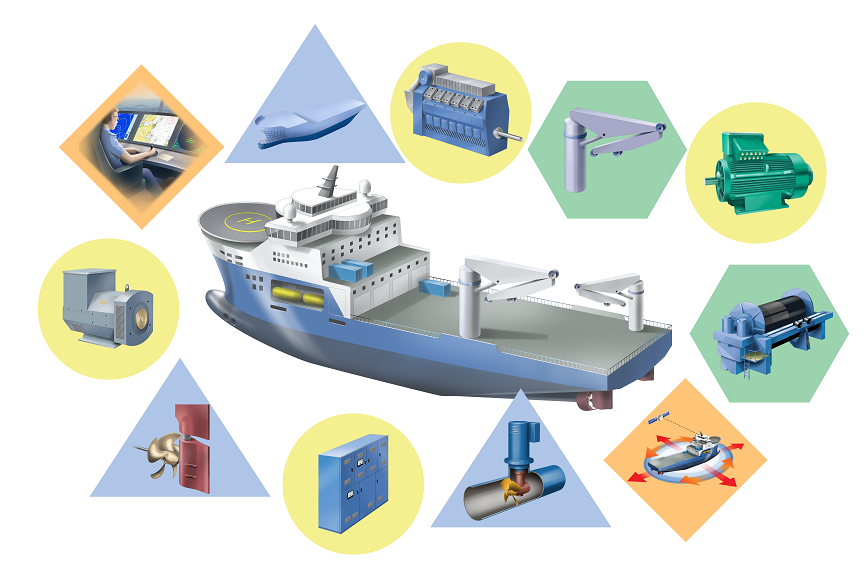}
	\caption{%
		Maritime systems and operations include a wide range of different engineering domains and physical systems with varying complexity and time scales.
		This naturally makes full-system simulation a challenging endeavor.
	}
	\label{fig:introduction_maritime_systems}
\end{figure}

If the optimization of the overall system performance is the goal, ship designs can, generally\footnote{%
Overall system performance can only be optimized in terms of the individual subsystems' performance if those subsystems are entirely independent of each other.
A simplification which will not hold for all but the most simple systems.
},
not simply be optimized for the performance of individual components and subsystems, but need to be optimized with respect to total operational performance.
Only with the interactions between components, the surroundings, control systems and software, and operators properly accounted for, is it possible to choose a system design with desirable characteristics---such as fuel efficiency, maneuverability and safety.
With more advanced operations requiring more power, interaction, and timing, system performance simulation will become even more important in the future.
Full-system simulation, however, remains a challenging and elusive task for at least two reasons:
\begin{enumerate}
	\item Typical maritime systems and operations are difficult and complex to model and simulate by nature:
	they are characterized by intricate interactions between a wide range of physical and engineering domains with dynamics taking place on vastly different time scales, see Fig.~\ref{fig:introduction_maritime_systems}.
	Compare, for example, the slow dynamics of large mechanical systems to the fast response of electronic components.
	\item This complexity and diversity is reflected in a simulation landscape that is riddled with specialized tools for different physical and engineering domains, with different interfaces and incompatible model representations.
Some existing simulation tools for maritime applications are highly advanced in terms of quality, functionality, and usability.
But they are mainly developed with research and the optimization of components and subsystems in mind, and lack interconnection capabilities.
General software solutions, on the other hand, are too inflexible, and offer model development and configuration that is too time-costly or inaccurate.
Moreover, the models themselves span a wide range of complexities and accuracies, including continuous as well as discrete behavior, and different focuses depending on the analyzed phenomena.
Consequently, understanding how different subsystems interact with each other and how they influence overall system behavior becomes all the more challenging.
\end{enumerate}

Traditionally, simulations of closely-coupled subsystems are constructed from the ground up, resulting in monolithic simulations for custom interfaces that are too application specific, too customized, and too costly in terms of development time.
The ability to assemble re-usable and interchangeable subsystem models into virtual prototypes in a \emph{plug and play} manner---regardless of the environment in which they are developed---should cut down on development times significantly and enable rapid innovation.
Undoubtedly, this would be a big step forward for the maritime industry.

The integration of multi-physics simulations, human behavior, and multiple parallel maritime operations has already been successfully demonstrated in operational training simulators.\cite{Fossen2011}
However, to date there are no universally agreed upon methods or standards supporting total systems integration and the analysis of operational performance.
The project \emph{Virtual Prototyping of Maritime Systems and Operations}\cite{VIPROMA-website,Hassani2016} (ViProMa) was initiated within the Norwegian maritime industrial cluster by independent research organizations, universities, and industry partners\footnote{%
MARINTEK, SINTEF Fisheries and Aquaculture, NTNU, VARD, Rolls-Royce Marine, and Det Norske Veritas.
}
with the goal of developing an open, standardized framework and architecture for system simulation and virtual prototyping as a new platform for product development and cooperation: the \emph{Virtual Prototyping Framework} (VPF).
This framework includes
\begin{itemize}
	\item guidelines for model coupling,
	\item high-level interfaces for coupling models from different engineering and physical domains,
	\item low-level interfaces for coupling models from different tools,
	\item a full-system simulation software,
	\item and example models and demonstrators.
\end{itemize}
It aims to make the communication between costumers, designers, and product developers more efficient throughout the design process.
It also facilitates the consistency and availability of objective \emph{Key Performance Indicators} (KPIs) if they are integrated into the prototyping system.

\subsection{Outline}
\label{subsec:introduction:outline}

In this paper, we discuss the development of a common technology platform and infrastructure supporting virtual prototyping and simulation-oriented work processes for maritime systems and operations.
The ViProMa project serves as an exemplary framework that allows us to have a closer look at the knowledge gaps and challenges with respect to the simulation of entire maritime systems and operations.
The aforementioned guidelines which are at the heart of the VPF are continuously summarized and emphasized on throughout the text.
While the context of the present work is maritime systems and operations, we would like to emphasize that the challenges discussed and the solutions presented here are broadly applicable.
They should, thus, also be of interest and value to people outside of the maritime sector.

First, we give a brief overview of the requirements for a virtual prototyping framework for the maritime industry in section~\ref{sec:requirements}.
In section~\ref{sec:cosimulation}, we discuss the development of a common architecture for system simulation (the VPF).
We review different simulation approaches and justify our choice of using co-simulation (simulator coupling) in ViProMa.
We then touch upon the formidable task of defining standardized domain model interfaces in section~\ref{sec:interfaces}, in order to establish a modular framework with high interoperability and re-usability with a focus on maritime applications.
Section~\ref{sec:construction} brings these concepts together in order to elaborate on the construction of full-system models and the underlying challenges.
We then discuss our decision to write our own in-house co-simulation software in section~\ref{sec:software}, where we also present the challenges faced and the progress made so far.
Finally, we provide a conclusion and share our thoughts on future developments in section~\ref{sec:conclusion}.


\section{Virtual Prototyping Framework Requirements}
\label{sec:requirements}

The requirements for a framework for virtual prototyping in the maritime industry are shaped by its most important use cases.
In terms of the ViProMa project, these are:
\begin{description}
	\item [Vessel design]
		Comparison and optimization of concepts with respect to fuel efficiency, capabilities, operabilities, availability, maintainability, and opportunity for future expansion.
		This also includes virtual sea trials, and the testing of control systems and station keeping abilities.
	\item [Crew training]
		Extending the possibilities towards higher realism and the inclusion of harsher environments, and increasing the awareness of vessel and equipment limitations.
	\item [Decision support]
		Choice of vessel for a specific voyage, and aligning ship capabilities with weather windows and equipment capabilities with specific operations.
\end{description}
Requirements are also dictated by the desired workflow that such a framework should facilitate:
After a model of a subsystem has been developed, it is connected to the full-system simulator.
A scenario is then selected, the full-system simulation run, and behavior and capabilities evaluated.

Within the ViProMa project, this lead to the following core requirements imposed on the VPF.
Note that while these are specific to the project and the maritime industry, most of them are highly relevant to a much broader area of application.
\begin{enumerate}
	\item There has to be support for distributed and cross-platform operation due to heterogeneous technologies, tools, and platforms, and the possibility for workload distribution.
	\item Human-in-the-Loop and Hardware-in-the-Loop, and, thus, real-time operation, have to be supported for integration with training simulators, dynamic positioning (DP) and other control systems, and various types of machinery and equipment.
	This also helps to save time in factory acceptance tests.
	\item The ability to use components as black boxes to protect intellectual property and sensitive information has to be implemented.
	\item The framework must be license-free with no restrictions on commercial use to prevent vendor lock-in, lower the barrier of use, and guarantee a widespread commitment.
	\item Increasingly rigid time constraints in the industry demand sufficient performance with regards to the overall prototyping process, as well as the simulations alone.
	In addition, strategies to achieve reasonable accuracy and stability of full-system simulations need to be established and implemented.
	\item The framework has to be sufficiently easy to use and provide well-defined interfaces in order for the industry to actually adopt and use it.
	This is also crucial in light of future development and maintenance of the framework.
	\item It should have a complete component database containing at least generic domain models of varying model complexity.
	Because final designs are often re-used to save money and time, the database should also be easily searchable.
\end{enumerate}
The following sections are devoted to addressing the challenges and possibilities in realizing these core requirements.
Due to limitations in funding and time, however, some of the requirements listed here were only partially or not at all fulfilled within the scope of the ViProMa project.
This will be commented on in Sec.~\ref{sec:conclusion}.


\section{Distributed Co-Simulation}
\label{sec:cosimulation}

In setting out to prototype and simulate a complex maritime system (such as a modern offshore supply vessel), and capture all its relevant dynamics and interactions, one quickly realizes that the traditional monolithic simulation approach is too inflexible, too costly, and too inefficient.
A model of an entire vessel usually takes a long time to get ready for simulation, if constructed from the ground up, and re-use is often prohibited when faced with similar problem sets due to the developed model being too customized and too application-specific.

It is clear then, that a modular approach in which models of the relevant subsystems are interconnected and simulated together is favorable to cut down on cost and development time.
This also means that the process of making changes to a subsystem should be effortless, in the sense that it does not require modifications of any other parts of the full-system model (i.e., the open-closed principle should be adhered to).
There are, generally, three methods to combine several models:
\begin{enumerate}
	\item \label{itm:modelstrategy1} The use of a common modeling language into which all models are translated for the purpose of the simulation,
	\item \label{itm:modelstrategy2} the exchange of models between tools to run a simulation in one of them (\emph{model exchange}),
	\item \label{itm:modelstrategy3} and \emph{co-simulation} (simulator coupling).
\end{enumerate}
Considering the requirements discussed in Sec.~\ref{sec:requirements}, the first two modeling approaches have several drawbacks:
\begin{itemize}
	\item General software solutions typically are too inflexible and offer model development and configuration that is too time-costly or inaccurate.
	\item In general, they neglect the availability of matured and specialized domain-specific analysis software for maritime applications.
	\item Users are reluctant to change modeling languages or tools, as doing so is a major undertaking in practice, and may render investments in tool chains and training worthless.
	\item The choices for a suitable common modeling language or tool are strongly limited if the development of an \emph{open} framework for systems simulation and virtual prototyping is the goal.
\end{itemize}
In addition, the use of a common modeling language sometimes means abandoning the black box approach that protects Intellectual Property Rights (IPRs) and sensitive information.
Model exchange, however, may play a vital role in future developments for full-system simulation, and we shall touch upon this briefly in Sec.~\ref{subsec:construction:tight-coupling}.

\begin{figure}[h!tb]
	\includegraphics[width=\graphicswidth]{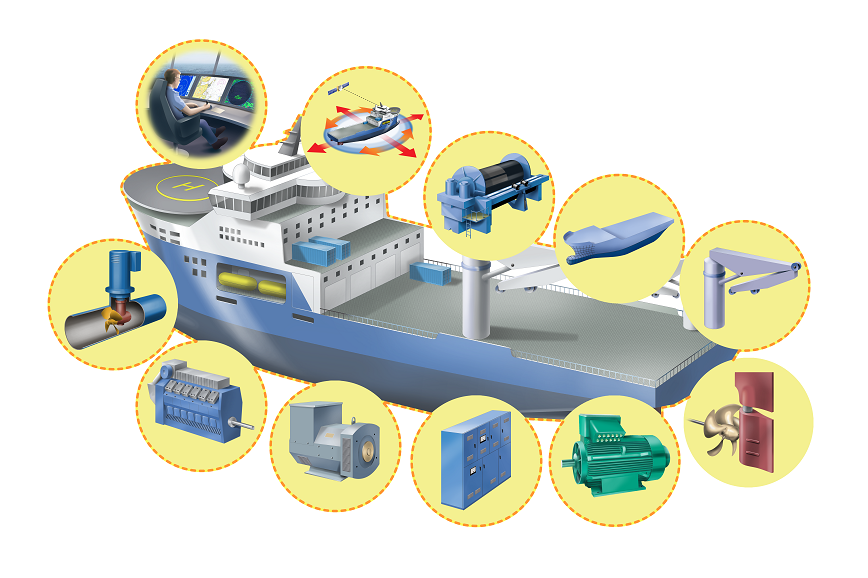}
	\caption{%
	In a co-simulation setting, different tools and models are interconnected and used independently and in parallel to form a full-system simulation.
	}
	\label{fig:tools_coupled}
\end{figure}

As mentioned previously, the fact that typical maritime systems and operations comprise of a wide range of physical and engineering domains naturally leads to a rather heterogeneous simulation landscape, with specialized tools and proprietary model representations.
While this may seem like a rather poor starting point for full-system simulation and virtual prototyping, it is precisely this modular structure of complex engineering systems, in conjunction with the availability of well-established domain- and application-tailored software, that lends itself quite well to co-simulation, see Fig.~\ref{fig:tools_coupled}.
The remainder of the present section is devoted to a discussion of the co-simulation approach and a brief review of existing co-simulation software and standards.

\subsection{Co-Simulation}
\label{subsec:cosimulation:cosimulation}

The basic idea behind co-simulation is the construction of systems from loosely coupled stand-alone models and the simulation across different subsystems.
Co-simulation facilitates the independent exchange and modification of components, and the use of the most suitable tools and solvers for any given subsystem.
This also extends to the possibility of separately taking care of initialization, pre-processing, time integration, and post-processing with different specialized tools.
This is very advantageous, because it allows for rapid and accurate model development that make efficient use of already available modeling software and languages without major investments in new tool chains or training.
Co-simulation further has the potential benefit of significantly reducing simulation time by using model-specific solvers and internal step sizes, and by allowing for the distribution of computational loads onto different computers or processor cores.
The possibility to conveniently hide internal dynamics and protect sensitive information is another attractive trait of simulator coupling, especially from an industrial perspective.

\begin{guideline}[Model coupling]
\label{guideline:Cosim}
	\guidelineCosim{}
\end{guideline}

Mathematically, co-simulation corresponds to the modular time integration of subsystems that are assumed to be independent in between discrete \emph{communication points} $t_i \in \{ t_0, t_1, \dots, t_N \}$.
The interactions between the subsystems are realized at these time points, and are expressed in the form of interface constraint equations,
\begin{equation}
\label{equ:cosimulation_connections}
	\vect{u}(t_i)
	=
	\vect{L}
	\vect{y}(t_i)
	,
\end{equation}
where $\vect{L}$ is a connection graph matrix relating the inputs $\vect{u}$ and the outputs $\vect{y}$.
This happens at a rate corresponding to a \emph{macro step size} $\Delta t_i$, such that $t_{i+1} =  t_i + \Delta t_i$.
In general, there is no guarantee that the pieces play together nicely, though.
Time synchronization and data exchange are important tasks, consequently, and sound and efficient communication between subsystems implies an adequate understanding of the architecture.

\begin{remark}
	Be cautious when selecting coupling method and co-simulation (macro) step size to avoid accuracy and stability issues.
\end{remark}

Because input variables are unknown to the subsystems during the time integration $t_i \rightarrow t_{i+1}$ and need to be approximated in general (and often held constant), co-simulation brings its own set of stability and accuracy issues.
Most commonly, this is remedied by selecting a sufficiently small macro step size.
At the same time, demands to keep the computational cost within reasonable bounds may, generally, require a lower limit, especially for real-time applications.

\begin{remark}
	For linear systems, the macro step size can be chosen from the eigenfrequencies, but it may be very difficult to find a good choice for nonlinear problems.
\end{remark}

In addition, there are several other subtleties and challenges to co-simulation accuracy and stability.
For example, the presence of algebraic loops can result in instability\cite{Kuebler2000}, and the presence of different time integration methods can actually decrease the overall accuracy of the full-system simulation to below the minimum accuracy of the individual subsystems\cite{Sicklinger2014Thesis}.
In general, the development of an efficient and robust co-simulation method that is easy to use and generally applicable is still ongoing research.\cite{Arnold2009}
Additionally, it is not always clear beforehand where to draw subsystem boundaries and how to choose a set of `good' interface constraint equations---both of which can play a significant role in determining stability and accuracy.

The ViProMa project tried to address many of these issues through original research:
\begin{enumerate}
	\item A novel \emph{Energy-Conservation-based Co-Simulation} method\cite{Sadjina2016,Sadjina2016b} (ECCO) was developed within ViProMa.
	It gives readily available feedback on global simulation quality, and significantly improves the accuracy and efficiency of non-iterative co-simulations.
	We shall discuss it briefly in Sec.~\ref{subsec:interfaces:high-level:residuals}.
	\item Additionally, an analysis tool for global stability in linear distributed dynamical systems has been proposed by combining dynamic stability and solver stability,\cite{Skjong2016b} both of which are intimately linked through local and global time steps.
	Under certain conditions, an algebraic solution of the total system can be constructed, and probed for global stability.
	However, this procedure can be very time consuming, and is only applicable for linear dynamical systems.
	Extensions to nonlinear dynamical systems have been studied as well, but the corresponding work is still ongoing.
\end{enumerate}
The task of establishing an efficient and robust general-purpose co-simulation methodology is far from completed, however.
Especially numerical stability is a nontrivial subject to study due to the inherent complexities (different solver methods of various orders, different coupling schemes of various orders, the presence of direct feed-through and algebraic loops, et cetera).

There exist non-iterative (explicit) and iterative (implicit) schemes to couple \sub{}s in a co-simulation, and time steps can be performed in parallel (Jacobi) or in serial (Gauss-Seidel).
The simplest and most straight-forward of all schemes is the explicit one with constant input approximation.
It is easiest to realize, keeps the exchange of coupling data to a minimum, and does not require the repetition of entire macro time steps (rollback).
Because of this, it is frequently used in industrial applications, and has also been the main focus for the ViProMa project so far.
It does, however, exacerbate the aforementioned stability and accuracy issues that co-simulation brings about naturally.

\subsection{Existing Co-Simulation Software and Standards}
\label{subsec:cosimulation:existing-software}

Among existing solutions for performing distributed co-simulations, the most prominent one is probably the \emph{High-Level Architecture} (HLA).
HLA is not one specific software package; rather, it is a standard which describes a general-purpose co-simulation architecture.
It was initially developed by the US Department of Defense for use in wargaming and training simulations, and was eventually made an IEEE standard.
The latest version of this is IEEE 1516-2010, commonly called \emph{HLA Evolved}.\cite{IEEE1516-2010}
Several HLA implementations exist today, both commercial and free.

Similar architectures include the \emph{Distributed Interactive Simulation}\cite{IEEE1278.1-2012} (DIS), which is the precursor of HLA and is even more geared towards military applications, and the \emph{Common Simulation Interface}\cite{Husteli2005} (CSI) developed by MARINTEK for the purpose of maritime vessel simulations.

These architectures are designed around the concept of a \emph{federation}, which is a group of independent subsystems (\emph{federates}) that communicate through a common \emph{Run-Time Infrastructure} (RTI).
The RTI is responsible for routing signals between the federates and for time synchronization.
The federates may be numerical simulations, hardware interfaces, human interfaces, et cetera.
Oft-stated advantages of HLA include \emph{interoperability}, in that federates may run on different platforms and use different simulation methods, and \emph{re-use}, in that federates used for one simulation may be easily re-used in another.
However, because the wire protocol between the federates and the RTI is not standardised, a federate created for one HLA implementation generally can't be used with a different implementation.
This, along with other reasons which will be explored in later sections, is why HLA was deemed unsuitable for the VPF.

Another standard we will mention here is the \emph{Functional Mock-up Interface} (FMI), which, unlike HLA, offers a way to make subsystems binary compatible with each other, thus, removing the need for recompilation and facilitating model sharing and co-simulation.
FMI has become a key component of the VPF, and we shall therefore describe it in more detail in section~\ref{subsec:interfaces:fmi}.



\section{Simulator Interfaces}
\label{sec:interfaces}

In order to ensure interoperability, modularity, and re-use between different models and simulators, well-defined interfaces are needed.
Such \emph{simulator interfaces} are a set of conventions that, if adhered to, allow a simulator to be coupled with other simulators using some co-simulation middleware.
Here, we distinguish between two levels of interfaces, which we shall refer to as \emph{high-level} and \emph{low-level} interfaces.

A high-level interface is concerned with the concepts which are being modeled and simulated.
That is, it deals with what a simulator represents on a physical level and the physical interpretation of data exchanged between simulators.
For example, it is crucial that the value of an output variable which represents a force in units of \si{\kilo\newton} in one simulator is not used for an input variable which represents a force in units of \si{\newton} in another---or one which represents, say, a voltage.
A high-level interface could prevent this by either mandating that certain quantities must have specific units, or by defining some mechanism whereby the units can be communicated, so that the co-simulation middleware can make the necessary value adjustments and/or prevent invalid connections.
On top of this, the interface can define \emph{groups} of variables which together have some physical significance.
An example of these are \emph{power bonds}, which are pairs of variables that represent different means of power transfer between entities.
Power bonds are discussed further in section~\ref{subsec:interfaces:high-level}.

Finally, at the highest level, one can define interfaces that represent categories of components or subsystems in the system being simulated.
For example, one could define that an `engine' has a power bond for rotational mechanical power and an output variable which represents fuel consumption.
Then, any simulator which adheres to this interface could be used to represent an engine, and could be replaced with any other simulator that has the same interface.
This opens great possibilities for `plug-and-play' construction of complex system models and, consequently, rapid evaluation and optimization of different designs.

The high-level interfaces are necessarily underpinned by one or more low-level interfaces, which are concerned with the finer details of how the co-simulation middleware interacts with the simulators.
At the lowest level, the physics involved are completely disregarded, and there is nothing preventing one from, for example, coupling a force variable with a voltage variable;
the interface deals in bits and bytes, not newton and volt.
One example of a low-level interface is what is known as an \emph{application binary interface} (ABI).
Among other things, an ABI defines how different types of data, such as integers, real numbers and textual data, are represented in computer memory.
Complementary to this is an \emph{application programming interface} (API), which specifies the names of program functions, which data they receive and return, and, to some extent, what the functions do.
For the VPF, the choice fell on the \emph{Functional Mock-up Interface}, which defines a simulator API and more.
This is discussed further in section~\ref{subsec:interfaces:fmi}.

\subsection{High-Level Interfaces}
\label{subsec:interfaces:high-level}

Power and energy are the universal currencies of physical systems.
Energy is conserved and continuous:
energy flows out of, or into, a system are always accounted for by appropriate energy storage and dissipation.
This is the theoretical foundation of bond graph theory\cite{Paynter1961,Breedveld1984}, which balances energy flows for each subsystem separately.
This way, they can be connected together in a modular fashion, while satisfying energy conservation and continuity for the entire system.
The energetic couplings between (sub)systems are realized with so-called \emph{power bonds}, which are defined by a pair of \emph{power variables}: a \emph{flow} and an \emph{effort}.
Their product is always a physical power---such as force and velocity, pressure and flow rate, or voltage and current.
The use of power bonds provides a complete and universal, energy-flow-centered connectivity between mathematical models of different engineering and physical domains.
As pointed out recently\cite{Sadjina2016}, they are thus perfectly suited for high-level interfaces for co-simulation.

\begin{guideline}[High-level interfaces]
\label{guideline:Powerbonds}
	\guidelinePowerbonds{}
\end{guideline}

\subsubsection{Power Bonds for Co-Simulation}
\label{subsec:interfaces:high-level:power bonds}

A power bond between two coupled simulators is realized by connecting two input--output pairs.
For example, a model of an electric generator could have a voltage as an output, which a connected electric consumer model would receive as an input.
In turn, the consumer would output an electric current, which the generator model accepts as an input.
These two exemplary models are then coupled via a power bond.

Demanding that physical couplings between simulators are realized through power bonds has a major advantage:
it allows to study the flow of energy between the subsystems directly using nothing but the simulator coupling values.
In fact, it makes it possible to directly observe if and where the conservation of energy is violated throughout a co-simulation, which, in turn, helps identify potential simulation issues, and provides valuable feedback about the quality of the results.

\subsubsection{Residual Energies and Energy Conservation}
\label{subsec:interfaces:high-level:residuals}

\begin{figure}[h!tb]
	\centering
	\def\svgwidth{\graphicswidth}
	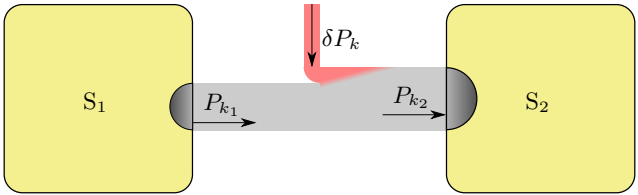
	\caption{%
		A residual power $\delta P_k = - (P_{k_1}+P_{k_2})$ emerges and distorts the dynamics of the full system when energy is exchanged between two \sub{}s, S$_1$ and S$_2$, in a co\hyp{}simulation
	}
	\label{fig:power_bond_residual}
\end{figure}

In general, energy is incorrectly transferred between two coupled simulators due to the fact that their states are evolved independently of each other between discrete communication time points, see Fig.~\ref{fig:power_bond_residual}.
In effect, energy is either created or destroyed through the co-simulation coupling during each macro time step.\cite{Benedikt2013}
This \emph{residual energy} directly alters the total energy of the overall coupled system.\cite{Sadjina2016}
It thus changes its dynamics, deteriorates simulation accuracy, and may pose a threat to numerical stability.

\begin{guideline}[Error estimation]
\label{guideline:Errorestimation}
	\guidelineErrorestimation{}
\end{guideline}

\begin{remark}
	The use of the \emph{Energy-Conservation-based Co-Simulation} method (ECCO) for reliable co-simulation error estimation is recommended.
	It is easily implemented, computationally inexpensive, and does not require the repetition of entire co-simulation step sizes.
\end{remark}

\begin{figure}[h!tb]
	\includegraphics[width=\graphicswidth]{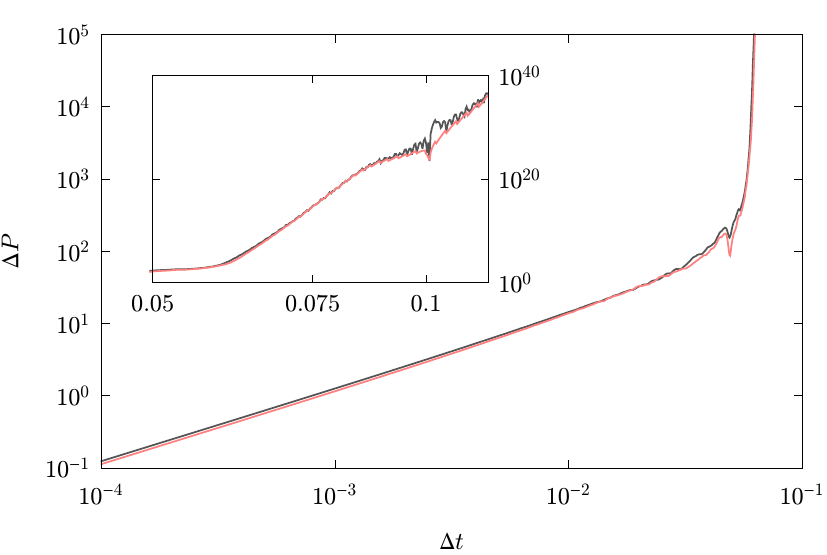}
	\caption{%
		Energy-conservation-based error estimation (red) compared to the actual error in the power $\Delta P$ (gray) as a function of the co-simulation step size $\Delta t$ for the benchmark model from Ref.~\onlinecite{Sadjina2016}.
		The critical step size is $\Delta t \approx \SI{0.059}{\second}$
	}
	\label{fig:ecco}
\end{figure}

If power bonds are used, such energy residuals are conveniently calculated from the coupling variable values alone.
These concepts are exploited by ECCO to obtain global error estimates.
Unlike virtually all other proposed co-simulation schemes, ECCO requires neither rollback nor any specific information on model implementations.
\footnote{%
	\label{footnote:Busch}
	To the best of our knowledge, the only other co-simulation method for error estimation and adaptive step size control with these qualities is the one described in Ref.~\cite{Busch2012}.
}
Consequently, it does not prohibit the use of commercial or legacy software (which often makes rollback inefficient or impossible), and it helps protect IPRs.
Because of this, it is especially attractive from an industrial perspective.
In addition, it accurately tracks coupling errors, even for relatively very large time steps, beyond which stability is already compromised, see Fig.~\ref{fig:ecco}.

\begin{remark}
	It is advisable to use methods which adaptively control the co-simulation step size.
	The use of ECCO for step size control is recommended to ensure approximately accurate energy transfers between \sub{}s.
\end{remark}

An adaptive control of the macro step size can easily be realized based on ECCO's error estimate.
This can improve the accuracy and efficiency of co-simulations substantially:
depending on system reticulation and \sub{}-internal solver accuracies, reductions in the error of \SI{93}{\percent} or more can be observed in benchmarks---at no additional computational cost.\cite{Sadjina2016,Sadjina2016b}
In practice, models may not support variable macro step sizes, however, generally leaving the potential for more accurate and more efficient co-simulations untouched.

\begin{remark}
	If power bonds are not applicable, other co-simulation methods for error estimation and adaptive step size control can be used, see Ref.~\onlinecite{Sadjina2016} and references therein.
	Almost all of these require the repetition of entire co-simulation step sizes and simulator-internal information, however.\footnotemark[\value{footnote}]
\end{remark}

\subsection{Functional Mock-up Interface}
\label{subsec:interfaces:fmi}

The \emph{Functional Mock-up Interface} (FMI) is a tool independent standard for the exchange of dynamic models and for co-simulation.\cite{Blochwitz2011}
The first version of the standard was published in 2010 as a result of the ITEA2 project MODELISAR.
Since 2011, maintenance and development of the standard have been performed by the Modelica Association, and a second version of the standard was released in 2014.\cite{Blochwitz2012}

FMI specifies that models should be packaged as \emph{functional mock-up units} (FMUs), which are archive files that contain model code for one or more platforms, along with metadata and documentation.
The standard defines the format and structure of files and directories in an FMU, as well as the APIs that must be implemented by the model code.
These APIs are defined in terms of the C programming language which, being the \emph{lingua franca} of programming, allows FMUs to be written in, and used from, practically any other language.

The FMI standard consists of two main parts:
\emph{FMI for Model Exchange} and \emph{FMI for Co-Simulation}.
FMI for Model Exchange specifies an interface for models that represent differential, algebraic, and discrete equations, which are typically coupled with and solved together with other models in some simulation software.
Important here is that the solver is supplied by the simulation software and is not part of the FMU code.
In contrast, FMI for Co-Simulation, which we shall focus on here, defines an interface for models which are bundled with their own solvers, and which can therefore be seen as separate simulators in themselves.
Several such `\sub{}s' will typically be coupled together in a co-simulation environment, a piece of software that enables data exchange between the \sub{}s, and keeps them synchronized in time.
All data exchange takes place at \emph{communication points} (sometimes called \emph{synchronization points}), between which each model is solved independently from the others by its own solver.
Note that we use the terms `model' and `simulator' freely here; in practice, there is nothing preventing these entities from being interfaces to hardware such as sensors, actuators, or devices for human input.

FMI for Co-Simulation is based on a master/slave model of communication and control, where \sub{}s are slaves that are controlled by a master algorithm.
The \sub{}s do not have any information about each other, nor about the simulation environment, except for the values they receive for their input variables.
Thus, they have no knowledge about or control over which other \sub{}s they are coupled to; the data is routed by the master algorithm.

The FMI standard is a fitting choice for maritime applications for three main reasons:
\begin{enumerate}
	\item It was created in collaboration with the automotive industry for many of the same reasons that we aim to design the VPF for the maritime industry.
	\item The standard is already supported by a large number of tools, for example by Dymola, JModelica.org, SIMPACK, SimulationX, and Simulink.
	\item FMI is completely open and free to use for any purpose.
\end{enumerate}
Finally, a decisive reason to choose FMI was that the alternative seemed to be to define a new interface and, to a large extent, reinvent the wheel.

\begin{guideline}[Low-level interfaces]
\label{guideline:FMI}
	\guidelineFMI{}
\end{guideline}

It should be emphasized at this point that FMI only specifies how the (co-)simulation software interacts with the models; it is \emph{not} in itself a simulation software, nor does it specify or restrict any other parts of the architecture of such a software.
More to the point, in a distributed co-simulation setting, FMI does not say how, or in what format, data are transported between the simulation nodes, nor how the nodes are time synchronized.
As such, FMI support can be implemented in almost any type of simulation software, and, indeed, the number of tools that support this standard is large and growing quickly.\cite{FMI-website}

\begin{remark}
	Whether or not any given FMU is compliant with the FMI standard can be checked with the free \emph{FMU Compliance Checker}.\cite{FMI-website}
\end{remark}

As an interesting side note, Awais \emph{et~al.}\ have proposed to use HLA as a co-simulation environment for FMI-based components.\cite{Awais2013}
This is despite the fact that HLA's federation model stands somewhat in contrast to FMI's master/slave structure:
federates are not passive slaves that simply wait to be given anonymous input data, but instead actively request, by name, the data they require.
The findings by Awais \emph{et~al.}\ appear to reflect this.
Their conclusion is that it is technically possible to use HLA as an FMI master, but to be able to make properly generic `FMU federates', the FMI standard must be extended so that the name mapping between federation data and FMU variables can be specified in the FMU metadata.

One aspect of FMI which should be mentioned here, as it has an impact on the model structures discussed in Sec.~\ref{sec:construction}, is the fact that FMUs are, for most practical purposes, \emph{closed for modification}.
That is, once an FMU has been created, there is no simple way to modify its behavior nor its external interface.
The model code is typically stored in compiled binary code form, and the numbers, types and names of input and output variables are fixed.
In some contexts, this puts severe limits on scalability.
For example, imagine that we have an FMU that models a body which is acted upon by external forces.
Then, we have to decide upfront, at model construction time, on a maximum number of force inputs.
If more forces are needed, the model source code must be modified and a new FMU compiled, or the force summation (and possibly coordinate transformations) must take place externally.

Also note that while FMI shows enormous potential and is continuously and actively refined, it also has various deficiencies.
Some aspects of co-simulation are altogether poorly addressed by the standard, and because limitations are often rather subtle, it is well worth pointing them out.
For example, it is possible to design FMUs and master algorithms that are standard compliant but exhibit nondeterministic and unexpected behavior.\cite{Broman2013}
It is also not directly clear how non-continuous models can be encoded as FMUs.\cite{Feldman2014,Tripakis2014,Broman2015}
Furthermore, FMI 1.0 has no general rollback mechanism beyond the previous macro time step, while FMI 2.0 makes rollback only optional, and error control is not addressed by FMI at all.\cite{Broman2013}



\section{Model Construction}
\label{sec:construction}

Let us now bring together the theoretical concepts and principles from the previous sections, and apply them to the construction of full-system models from stand-alone components and submodels.
We shall first elaborate on some general considerations and common challenges, before we reap the fruits of our efforts and move on to discuss some model examples at the end of the present section.

\subsection{\Sub{}s}
\label{subsec:construction:subs}

\Sub{}s represent the various components which are connected together to form a full-system simulator.
It is important to note that this concept does not only include time-dependent physical systems.
The actual functionality of a \sub{} is at the complete discretion of its developer, and includes:
\begin{itemize}
  \item mathematical models of physical systems evolving over time,
  \item control system implementations,
  \item scenario controllers (units which control and steer a simulation, e.g., by changing parameters during runtime),
  \item hardware interfaces (e.g.\ to allow direct hardware input to the simulation),
  \item and bridging to other software or co-simulations.
\end{itemize}
This list is not exhaustive by any means, and several other use cases are possible.
That some of these may warrant the use of a different concept will be the subject of Section~\ref{subsec:construction:function_units}.
But first, let us discuss the boundaries between \sub{}s.

\subsection{System Boundaries}
\label{subsec:construction:system_boundaries}

\begin{figure*}[h!tb]
  \includegraphics[width=0.9\graphicswidthfull]{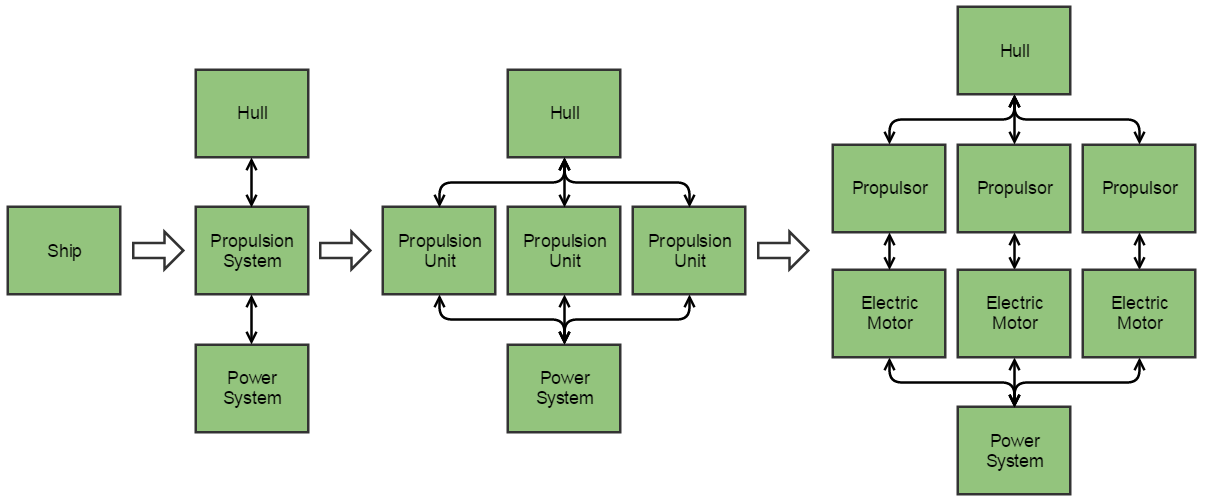}
  \caption{%
  	Different levels of modeled modularity of systems on board a ship
  }
  \label{fig:different-modularity-levels}
\end{figure*}

When modeling a complex system composed from many different subsystems in a co-simulation environment, one of the many challenges is to determine where to draw the boundaries between the separate modules.
To exemplify this, consider simulating a ship which consists of a power system, propulsion units, and a hull.
A few examples of how to divide such a system into different subsystems are shown in Fig.~\ref{fig:different-modularity-levels}:
One option is to include everything in one single \sub{}, but this makes changes difficult or impossible to perform for users not having access to the \sub{} implementation.
By splitting the total system into several parts the modularity will increase, but so will the complexity.

\begin{figure}[h!tb]
  \includegraphics[width=.6\graphicswidth]{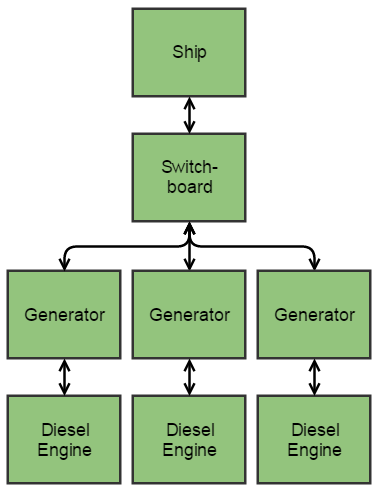}
  \caption{%
  	Example of system modularization for a ship model with a special focus on power-system dynamics
  }
  \label{fig:power-system-modularity}
\end{figure}

Where to draw the line is very much up to the designer, but it should be motivated by what is under investigation.
Typically, a higher level of modularity makes it easier to include individual models of high fidelity.
Therefore, a seemingly good rule-of-thumb is to keep modularity and fidelity high in parts of the system that are of interest, while lowering them for the rest of the system.
This is exemplified by Fig.~\ref{fig:power-system-modularity} which shows a model with high modularity in the power system.
The entire remainder of the ship is captured inside only one \sub{}, by contrast, and its only purpose is to provide `good-enough' dynamics to give studies of phenomena in the power system the necessary embedding.
Keep in mind, however, that there needs to be a trade-off between this modeling modularity on one hand, and accuracy and stability on the other:
as discussed in Sec.~\ref{subsec:cosimulation:cosimulation}, higher modularity may, generally, \emph{decrease} the overall accuracy for the part of the system under investigation, and therefore, for the entire co-simulation.

\begin{guideline}[System reticulation]
\label{guideline:Reticulation}
	\guidelineReticulation{}
\end{guideline}

\begin{remark}
	The way a given system is split up for co-simulation generally affects accuracy and stability.
	An ill-chosen system reticulation can cause a significant deterioration of the quality of the co-simulation results, or even an entirely unstable simulation. 
\end{remark}

As an example, consider the two different system reticulations for the simple linear quarter car benchmark model studied in Ref.~\onlinecite{Sadjina2016}:
the mean global error increases by almost a factor of ten when choosing the less favorable reticulation, and relatively large (constant) macro step sizes, which cause no issues with one reticulation, lead to instability with the other.
Another practical example is given by the development of a Dynamic Positioning (DP) controller for ViProMa.
The two main parts of such a controller are a high-level motion controller and a \emph{Thrust Allocation Algorithm} (TAA).
The initial DP controller design kept the high-level motion controller and the TAA in two separate \sub{}s.
This seemed like a good idea from a modularity perspective, because it allowed to easily swap out or modify either of the two independently.
However, experience showed that the additional macro time step delay introduced by splitting the DP controller this way made the two parts much more difficult to tune, and fragile to changes in the macro time step size.
Putting both, the high-level motion controller and the TAA, inside one \sub{} greatly improved the tunability and robustness.

\subsection{Tight Coupling}
\label{subsec:construction:tight-coupling}

Another important issue, which kept resurfacing when establishing the VPF, is how to sensibly deal with tightly-coupled subsystems.
Consider, for example, the rigid mechanical connection between a vessel's hull and crane.
Ideally, such a connection calls for solving the hull--crane system as one, and the straight-forward (explicit) co-simulation of both as separate \sub{}s with separate solvers is unfeasible.
In other words, it may be best to simply refrain from splitting tightly-coupled systems for co-simulation altogether.

\begin{guideline}[Tight coupling]
\label{guideline:Tightcoupling}
	\guidelineTightcoupling{}
\end{guideline}

This, however, challenges the general and modular virtual prototyping and full-system simulation approach that is so desirable from a practical point of view.
It may still be possible to allow for sufficient flexibility in a large number of applications if generic models are used that offer sufficient parameterization.
But in some cases it may simply be impractical to avoid exposing tight couplings between \sub{}s, and several strategies have been proposed as solutions:
\begin{enumerate}
	\item The most common approach is to include dampers and springs in one of the models.
	This works fine in practice, but has some notable drawbacks: Firstly, suitable parameters need to be found to configure these additional elements. This, secondly, means that the relatively stiff springs and strong dampers required introduce relatively small time constants, which can easily become a challenge for co-simulation.
	\item A possibly very fruitful approach is to implement tightly-coupled subsystems using \emph{model exchange}.
	Then, the models can be shared and modularly integrated, while they are being solved jointly by one solver.
	While this generally still satisfies the black box criterion, it may mean abandoning well-established simulation tools for some users.
	It could also proof difficult to find appropriate \emph{open} software solutions enabling the use of model exchange in a co-simulation environment.
	\item Yet another way of dealing with rigid coupling is to use advanced iterative co-simulation approaches utilizing \sub{} outputs along with their respective Jacobians.
	An example of this is the \emph{Interface Jacobian-based Co-Simulation Algorithm}\cite{Sicklinger2014}, in which coupling conditions are solved iteratively with Newton's method.
	Unfortunately, such approaches are presently mainly of academic interest:
	generally requiring Jacobians from \sub{}s---along with their ability to redo macro time steps---is simply an unrealistic condition short-term and medium-term.
	\footnote{%
		In the long run, however, such co-simulation techniques are very promising, and will hopefully become the status quo.
	}
	\item Finally, one can neglect the effects of one of the subsystem on the other.
	For example, a hull would simply dictate the position of a crane attached to it, while the forces sent back to the hull are only due to the crane's inertia.
	This approach is only sufficient, however, if an accurate representation of the forces is not desired.
\end{enumerate}
As reflected by these choices, tight coupling is a complex and sensitive issue for virtual prototyping and co-simulation, and requires further research.
For the time being, the best course of action will be determined by the case at hand, and can hopefully be found with the examples given here.

\subsection{Connections}
\label{subsec:construction:connections}

Connections define the interactions between \sub{}s and are intimately interrelated with the system reticulation.
Therefore, they need to be treated and implemented carefully.
Mathematically, connections are expressed via the connection graph matrix $\vect{L}$ of Eq.~\eqref{equ:cosimulation_connections}, but in a co-simulation they are only enforced at the discrete communication time instances.
This naturally creates challenges for accuracy and stability, as discussed in Sec.~\ref{subsec:cosimulation:cosimulation}.

As we have just seen, system boundaries need to be chosen sensibly, as to not compromise modularity, simplicity, accuracy, and numerical stability.
At the same time, factual interactions between parts of a system and with the environment need to be captured accurately, given their relevance to the use case at hand, available computational capabilities, and modeling abilities and knowledge.
Physical connections between subsystems are implemented as high-level interfaces between \sub{}s, as discussed previously in Sec.~\ref{subsec:interfaces:high-level}.
In doing so, ensuring the compatibility and correct wiring between the corresponding \sub{} inputs and outputs is of great importance.
Let us elaborate on this point in the next two sections.

\subsection{Function Units}
\label{subsec:construction:function_units}

In order to connect the output of one \sub{} to the input of another, it may be necessary to apply some simple transformation in between.
For example, in order to match an output's unit to that of an input, a conversion from \si{\radian/\second} to \si{rpm} could be needed, or a simple summation, to satisfy Kirchhoff's circuit laws or compute the net force acting on a body (see Fig.~\ref{fig:hydrodynamic-models}).
Another example includes cases where the number of connections for a given \sub{} is variable or unknown.

In order to maintain modularity and ease-of-use, such transformations can not take place inside the \sub{}s, nor should they be taken care of by a master or hard-coded into the signal routing.
One solution is to deploy a \sub{} between the others, dedicated to translating its input into an appropriate output form.
While this does work, it has the unfortunate drawback in that it leads to an additional time delay between the original output and input the length of one macro time step.

To avoid this, the theoretical concept of so-called \emph{Function Units} (FUs) was introduced.
These act very much like \sub{}s, but they are \emph{time independent} and perform their calculations \emph{in between} macro time steps, at communication points.
Transformations between inputs and outputs are, thus, carried out efficiently between time steps, can be easily applied in a modular and flexible fashion, and do not introduce unwanted time delays.
In addition, Function Units are also well suited to represent dynamics on timescales that are much shorter than the ones that are relevant for the problem being studied.
For example, in a simulation of a ship and its propulsion machinery, many electrical infrastructure components, such as switchboard breakers, would be ideal candidates for Function Units.

\begin{guideline}[Function Units]
\label{guideline:FUs}
	\guidelineFUs{}
\end{guideline}

Mathematically, this concept corresponds to a generalization of the connection graph matrix of Eq.~\eqref{equ:cosimulation_connections}, such that
\begin{equation}
\label{equ:connectiongraph_general}
	\vect{u}(t_i)
	=
	\vect{L}
	\big(
		\vect{y}(t_i)
		,
		t_i
	\big)
	,
\end{equation}
where $\vect{L}$ is now, in general, a function of the outputs $\vect{y}$ and the time.
Common linear transformation---such as unit conversions, summations, and coordinate transformations---may still be simply expressed in the form given by Eq.~\eqref{equ:cosimulation_connections}, however.

It needs to be emphasized at this point that the concept of Function Units is mainly a theoretical one, and is still awaiting proper practical verification.
Yet, their versatility and independence on time makes them essentially indispensable for constructing complex full-system models from re-usable stand-alone \sub{}s:
virtually any type of code can be put inside an FU, and changes can quickly and easily be made on the system level \emph{after} all the subsystem modeling is done.

\subsection{Hybrid Causality}
\label{subsec:construction:hybrid_causality}

The computational causality of a subsystem also plays an important role in defining the connectivity with other subsystems.
In some cases, the causality and, thereby, the connectivity of a \sub{} are difficult to determine beforehand without knowledge of the connecting environment.

To illustrate this causality--connectivity dependency, consider a simple mass--damper--spring system.
Using integral causality, it is given by
\begin{equation}
\label{equ:massdamperspring_integral}
\begin{split}
	\dot{x}_1
	&=
	x_2,
	\\
	\dot{x}_2
	&=
	\frac{1}{m}
	(
		\tau- d x_2 - k x_1
	)
	,
\end{split}
\end{equation}
where $x_1$ and $x_2$ are the position and the velocity of the system, respectively, $m$ is the mass, $d$ is the damping coefficient, $k$ is the spring stiffness, and $\tau$ is an excitation force given to the system as an input.
If power bonds are used to connect to other \sub{}s in accordance to Guideline \ref{guideline:Powerbonds}, the output of the mass--damper--spring system is given by $x_2$.
But Eq.~\eqref{equ:massdamperspring_integral} can also be written using differential causality, such that
\begin{equation}
\label{equ:massdamperspring_differential}
\begin{split}
	\dot{x}_1
	&=
	v,
	\\
	\tau
	&=
	m \frac{\diff v}{\diff t}
	+
	d v
	+
	k x_1
	.
\end{split}
\end{equation}
Input and output have been switched, with the velocity $v$ now constituting the input, while the output is given by the force $\tau$.
Both these causality options, and, thereby, connectivity options, may be equally relevant, depending on the \sub{} environment in the full system.
Note that one state was lost from switching from integral to differential causality.

Different strategies exist to resolve conflicts in computationally causality between \sub{}s:
The model in question can simply be implemented with a specific causality and connectivity.
This is a quick and easy fix, but it is dependent on the full system specifics, and not very attractive from a modularity point of view.
Alternatively, the interface constraint equations~\eqref{equ:cosimulation_connections} can be solved iteratively.
As mentioned in Sec.~\ref{subsec:cosimulation:cosimulation}, it is often preferred to avoid the repetition of entire co-simulation time steps, however, and employ explicit co-simulation schemes instead.
This leaves us with the final option of implementing the \sub{} as a \emph{hybrid causality model}, that is, with the possibility to switch between different causality options---in some cases even on-line during a simulation.

\begin{guideline}[Hybrid causality]
\label{guideline:Causality}
	\guidelineCausality{}
\end{guideline}

\begin{remark}
	There are a few subtleties when switching causality on-line during a simulation:
	For one, the state space changes, which either requires a solver that is suited for both systems, or the switching of the solver along with the model.
	Secondly, initial conditions need to be chosen carefully in order to not violate the conservation of energy, and to avoid discontinuities in the input and output signals.
\end{remark}

Hybrid causality models can be implemented in several different ways.
For example, two models with different causality options can be implemented in parallel as one \sub{} and switched between.
The switching itself can depend on logical choices, and active and passive inputs and outputs.
Note that the submodel with differential causality may need to be solved by a more advanced numerical solver.
A different approach\cite{Skjong2016b} is the use of a low-pass filter---effectively acting as an integrator---to regain the state that is lost when switching from integral to differential causality, as in the case exemplified by Eqs.~\eqref{equ:massdamperspring_integral} and \eqref{equ:massdamperspring_differential}.
This method keeps the number of states in the system constant and removes any need for iterations.

A typical application of hybrid causality models within the realm of maritime systems is the modeling of weak marine power grids containing more than one energy source (such as generators, batteries, or fuel cells).
Because only one source can determine the power grid voltage, the sources need to be implemented as hybrid causality models, in general.
This is especially true if they are to be added and removed from the grid during the simulation.\cite{Skjong2017}

\subsection{Model Examples}
\label{subsec:construction:model_examples}

Now that we have set the stage, let us demonstrate the application of the concepts discussed so far by use of two examples from the ViProMa project.

\subsubsection{Vessel Model}
\label{subsec:construction:model_examples:vessel}

\begin{figure}[h!tb]
  \includegraphics[width=\graphicswidth]{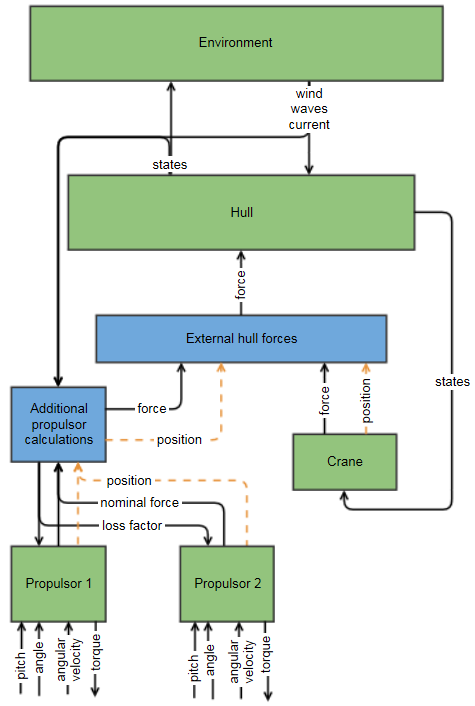}
  \caption{%
  	Illustration showing how propulsors, crane, trawl, and environment \sub{}s have been connected for ViProMa.
  	The system includes \sub{}s (green) as well as Function Units (blue).
  }
  \label{fig:hydrodynamic-models}
\end{figure}

First, consider the system shown in Fig.~\ref{fig:hydrodynamic-models}, with a hull model connected to propulsor models and a model of the environment.
In addition, a crane model is shown to illustrate the fact that there may be multiple subsystems acting on the hull with a significant force.
All of these are realized as \sub{}s (green).
Additionally, FUs (blue) take care of some simple signal algebra and additional time independent calculations.

The power bonds connecting the crane and trawl models to the hull are realized as \emph{force--velocity} variable pairs, where the velocity is a part of the hull output \emph{states}.
The propulsor models here are RPM controlled, and so the variable pairs constituting the corresponding power bonds are of the type \emph{torque--angular velocity}.
They are thought to be connected to an actuator which drives them (e.g.\ el-motor or diesel engine, not shown).
All propulsor \sub{}s are also connected to an FU labeled \emph{Additional propulsor calculations}.
Its function is to sum up all the forces, and to perform additional calculations to determine the thrust loss for each propulsor individually.
Another FU, labeled \emph{External hull forces}, sums up all the forces acting on the hull, and applies the resulting net force to its center of gravity.
In order to do so, it also requires knowledge about the various points of attack for the forces, which it receives via the additional \emph{position} signals.
All of these calculations are time independent and, thus, are conveniently put inside of Function Units.
Doing so also brings about all the benefits of FUs discussed in Sec.~\ref{subsec:construction:function_units}:
a flexible number of inputs and outputs (to connect any number of \sub{}s), the possibility to quickly modify the signal algebra (to alter the thrust loss calculations, for example), and (quasi-)instantaneous execution without any delay in logical time.

The current state of environment modeling---specifically waves---in the ViProMa project is that principal data like wave height, wave period, and wave spread are specified for each dependent \sub{} individually.
In the present example, this is the case for the \emph{Hull} and \emph{Additional propulsor calculations} \sub{}s, both of which use the same underlying implementation to realize the wave spectrum from that data.
In the future, we would like to centralize wave spectrum construction in one location, and pass it on to \sub{}s as required.
This is challenging, however, because the amount of data for wave realization can be as large as \num{16384} wave components.

\subsubsection{Power Plant}
\label{subsec:construction:model_examples:power_plant}

\begin{figure}[h!tb]
  \includegraphics[width=0.6\graphicswidth]{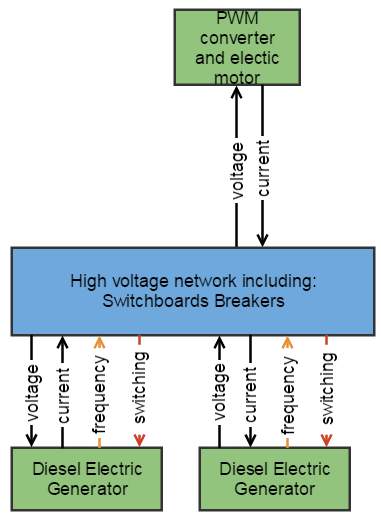}
  \caption{%
  	Illustration showing how a diesel electric generator, a switchboard, and an electric motor \sub{} have been connected in ViProMa.
  	The system includes \sub{}s (green) as well as Function Units (blue).
  }
  \label{fig:power-system}
\end{figure}

An example of a power plant model which connects \emph{Diesel Electric Generator}, \emph{Switchboard}, and \emph{EL-motor} models is shown in Fig.~\ref{fig:power-system}.
The \emph{frequency} signals between the diesel electric generator models and the switchboard are necessary because these specific models are using the direct-quadrature-zero transformation (dq0), and the \emph{switching} signals are used to swap the (computational) causality of the models.
As explained in Sec.~\ref{subsec:construction:hybrid_causality}, the latter is needed because only one generator model can set the voltage for the switchboard when they are coupled in the manner depicted.
The FU in this example represents a switchboard which acts as a summer/splitter for the voltage and current, and also captures breaker functionality.
In some cases, a physical model of a breaker may be needed and a \sub{} deployed instead, but the breaker dynamics are commonly quasi-instantaneous and have no effect on the rest of the system.

\section{Simulation Software}
\label{sec:software}

To support and demonstrate the use of the Virtual Prototyping Framework, the ViProMa project has developed \emph{\cosimSW{}}, a co-simulation software built from the ground up with FMI support and all the requirements described in section~\ref{sec:requirements} in mind.
Being designed for FMI, it has the same master/slave structure that we described briefly in section~\ref{subsec:interfaces:fmi}.

This software has two primary responsibilities:
The first is \emph{communication}, in that it transports data between \sub{}s, possibly over a network, ensuring that output values are routed to the correct input variables.
The second is \emph{synchronization}, in that it issues commands to all the \sub{}s that tell them when to perform a new time step, and how far to simulate before reaching the next communication point.

Since \cosimSW{} supports distributing simulations over a network, it becomes necessary to actually start the slave programs on each computer when a new simulation is run.
This is handled by a small server program called a \emph{slave provider}, which runs on each of the machines that are set up to participate in simulations.
This program is responsible for loading the FMUs available on that machine, publishing information about them on the network, and spawning slaves at the request of a master.
An FMU is a bit like a \emph{class} in object-oriented programming terminology, in that it represents a `blueprint' for a model, and several instances of that model---slaves, in other words---can usually be created from one FMU.
Each such instance then typically has its own state, as well as its own inputs and outputs.

Currently, the \cosimSW{} slaves simply log their own simulation results directly to file.
However, in the near future, a new type of simulation entity called \emph{observer} will be added to the system.
Unlike slaves, observers are privy to a lot of information about the structure and state of the simulation: which units are on-line, their inputs and outputs, and so on.
Observers have no output values; in fact they have no way of affecting the course of the simulation at all.
Typical examples of systems which could be implemented as observers include visualization systems and data loggers.

\begin{figure*}[h!tb]
  \includegraphics[width=0.77\graphicswidthfull]{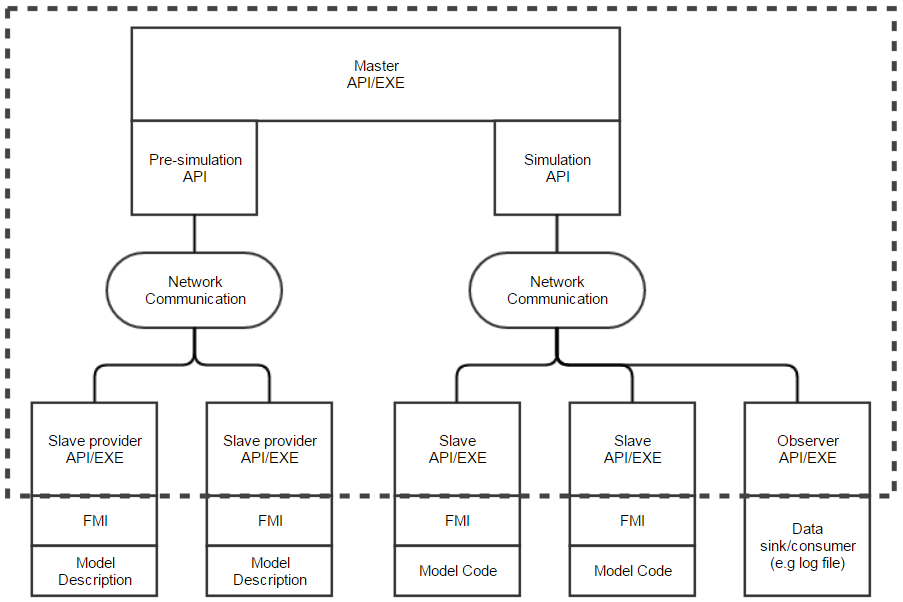}
  \caption{%
    A diagram that shows the various components in a \cosimSW{} simulation.
    Everything inside the dashed rectangle is formally part of \cosimSW{}.
    By \emph{API/EXE} we mean that the functionality is offered both in the form of a C++ programming interface and as a ready-made executable application.
  }
  \label{fig:coral-structure}
\end{figure*}

\cosimSW{} is implemented as a software library for the C++ programming language, so it can be embedded in programs that need to perform co-simulations.
It also comes with a set of command-line tools that allow users to run simple simulations which are configured via text files, and which double as examples that demonstrate how the C++ APIs may be used.
See Fig.\ \ref{fig:coral-structure} for an overview of the software structure.

One may rightfully ask why it was deemed advantageous to create a new co-simulation software from scratch, rather than use an existing one, such as for example one of the many HLA implementations.
The answer is that we were unable to find an existing software that fulfilled all the requirements described in section~\ref{sec:requirements} to our satisfaction.
In addition, a goal of the ViProMa project was to research and develop novel simulation methods and technologies, and in this respect it is very useful to start with a blank slate and to have a code base which is under our full control.
The entire software will be released under a permissive open-source license when the ViProMa project is complete.

\section{Conclusion}
\label{sec:conclusion}

While it is widely recognized that new ship designs need to be optimized with respect to the overall operational performance---rather than the performance of components and subsystems---the simulation landscape is riddled with specialized tools for different physical and engineering domains and incompatible model representations.
General software solutions, on the other hand, often fall short of providing the flexibility, efficiency, and accuracy needed for model construction in an increasingly competitive environment.
To date, there are no universally adopted standards and tools supporting total systems integration and the analysis of operational performance.

In the present paper, we discussed the development of an open and standardized technology platform and infrastructure for virtual prototyping and full-system simulation for maritime systems and operations: the \emph{Virtual Prototyping Framework} (VPF).
Its vision is to facilitate the rapid development and sharing of subsystem and component models, and their joint simulation to assess full-system performance, optimize and verify designs, and establish a new arena for collaboration in the maritime industry.
The biggest knowledge gaps and challenges were exposed, and solution strategies offered, with a focus on high interoperability, modularity, and re-usability.
The application of the VPF, its guidelines, and its interfaces was demonstrated by use of model examples from the \emph{Virtual Prototyping of Maritime Systems and Operations}\cite{VIPROMA-website} (ViProMa) project.
Our in-house co-simulation software \emph{Coral} was introduced, and original research towards establishing efficient, accurate, and stable co-simulation methods was discussed.

At the heart of the VPF lies a set of guidelines for model coupling, describing best practices for full-system model construction and simulation.
In brief, these are:
\begin{description}
\item[Guideline~\ref{guideline:Cosim}]
	\guidelineCosimshort{}
\item[Guideline~\ref{guideline:Powerbonds}]
	\guidelinePowerbondsshort{}
\item[Guideline~\ref{guideline:Errorestimation}]
	\guidelineErrorestimationshort{}
\item[Guideline~\ref{guideline:FMI}]
	\guidelineFMIshort{}
\item[Guideline~\ref{guideline:Reticulation}]
	\guidelineReticulationshort{}
\item[Guideline~\ref{guideline:Tightcoupling}]
	\guidelineTightcouplingshort{}
\item[Guideline~\ref{guideline:FUs}]
	\guidelineFUsshort{}
\item[Guideline~\ref{guideline:Causality}]
	\guidelineCausalityshort{}
\end{description}
These guidelines are by no means meant to be complete or even final.
But their introduction into, and adaption by, the maritime industry would be a major step ahead towards finding tomorrow's design and technology solutions.
Note, however, that these guidelines are not specific to the maritime domain and should apply to a wide range of engineering and scientific applications.

The definition of guidelines was motivated by a set of core requirements, which were derived from the most important use cases---vessel design, crew training, and decision support---and the desired work flow for the prototyping framework.
As mentioned in Sec.~\ref{sec:requirements}, some of these had to be left to future research due to insufficient funding and time.
Most notably, we fell short of implementing the following three aspects:
\begin{enumerate}
	\item A basic open library layer of generic and, to a certain extent, parameterized domain models and components should be available for the VPF to configure ships and operations for simulation and optimization studies.
	This is also important for demonstration purposes, and to aid in the wide-spread acceptance and adaption of the framework.
	Creating such a database is an extensive undertaking, however.
	\item Only some testing and research\cite{Skjong2017b} was done with respect to real-time capabilities.
	While, in principle, there is nothing preventing the use of real-time simulations with the VPF as introduced in the present work, more research and verification is surely needed.
	At any rate, there are a few challenges that need to be considered:
	For example, computational efficiency can easily become an issue, because real-time \sub{}s (such as hardware interfaces) will usually dictate a lower limit for the co-simulation step size.
	Moreover, network latency effects and noise in measurement signals have to be dealt with.
	These points can, naturally, also further exacerbate the co-simulation-inherent accuracy and stability issues discussed in Sec.~\ref{subsec:cosimulation:cosimulation}.
	A promising project is the \emph{Advanced Co-Simulation Open System Architecture}\cite{ACOSAR-website} (ACOSAR), which aims to supplement the FMI standard with a global industry standard for real-time system integration and co-simulation.
	\item One important aspect of specifying high-level simulator interfaces is the definition of standardized categories which represent various component and subsystem groups---such as engines or cranes---along with standardized model properties.
	This would facilitate rapid prototyping in a convenient `plug-and-play' fashion, as mentioned in Sec.~\ref{sec:interfaces}.
\end{enumerate}
Hopefully, the present work can contribute to the advancement of virtual prototyping and flexible full-system simulation, provide a basis for further research, and have its aforementioned shortcomings amended through future work.


\begin{acknowledgements}

This work was funded by the Research Council of Norway (Grant Number 225322), and the industrial partners in the ViProMa project consortium (VARD, Rolls-Royce Marine, and DNV GL).
We are grateful for their financial support.

\end{acknowledgements}


\nocite{*}    
\bibliographystyle{plain}
\bibliography{wp1}


\end{document}